\documentclass[twocolumn, twoside, slac_two]{revtex4}
\usepackage{graphicx}
\usepackage{fancyhdr}
\usepackage{amsmath}
\usepackage{amssymb}
\newcommand{\GeV}{\ \text{GeV}}

\newcommand{\cm}{\ \text{cm}}

\newcommand{\MP}{M_{\text{P}}}
\newcommand{\GF}{G_{\text{F}}}
\newcommand{\mG}{m_{3/2}}
\newcommand{\mN}{m_{\chi_1^0}}

\begin{document}

\title{Galaxy Clusters and Gamma-Ray Lines:\\Probing Gravitino Dark Matter with
the Fermi LAT}

%

\author{Xiaoyuan Huang}
\affiliation{National Astronomical Observatories, Chinese Academy of Sciences,
Beijing, 100012, China}
\author{Gilles Vertongen}
\affiliation{Institut d'Astrophysique de Paris, UMR-7095 du CNRS, 98 bis bd
Arago, 75014 Paris, France}
\author{Christoph Weniger}
\affiliation{Max-Planck-Institut f\"ur Physik, F\"ohringer Ring 6, 80805
M\"unchen, Germany}

%

\begin{abstract}
  If dark matter particles are not perfectly stable, their decay products
  might be seen in the cosmic-ray fluxes. A natural candidate for decaying
  dark matter is the gravitino in R-parity violating scenarios. In the
  relevant GeV-TeV energy range, the Fermi Large Area Telescope (LAT) is now
  measuring cosmic gamma-ray fluxes with an unprecedented precision. We use
  the Fermi LAT gamma-ray data to search for signatures from gravitino dark
  matter particles, concentrating on gamma-ray lines and galaxy cluster
  observations. Implications of our results for the decay length of the
  next-to-lightest superparticle, which could be seen at the LHC in the near
  future, are discussed.
\end{abstract}

\maketitle

\thispagestyle{fancy}

\section{Introduction} A theoretically well motivated example for
\textit{decaying dark matter} is the gravitino $\psi_{3/2}$, which appears in
locally supersymmetric extensions of the Standard Model. In scenarios where
$R$-parity is mildly violated and the gravitino is the lightest superparticle
(LSP), thermal leptogenesis, gravitino dark matter and primordial
nucleosynthesis are naturally consistent~\cite{Buchmuller:2007ui}.  Within
this framework, the gravitino would decay with cosmological
lifetimes~\cite{Takayama:2000uz}, making its decay products potentially
observable in the cosmic-ray fluxes~\cite{Bertone:2007aw, Ibarra:2007wg,
Ishiwata:2008tp, Choi:2009ng, Choi:2010jt}. For gravitino masses
$m_{3/2}\lesssim 200\GeV$, the most prominent feature in the decay spectrum is
an intense gamma-ray line, produced by two-body decay into neutrinos and
photons, $\psi_{3/2}\to\gamma\nu$~\cite{Ibarra:2007wg}; this line can be
searched for in the gamma-ray fluxes observed at high latitudes.  For larger
gravitino masses $m_{3/2}\gtrsim 200\GeV$, the branching ratio into gamma-ray
lines is suppressed, and instead the decay modes $\psi_{3/2}\to W^\pm
\ell^\mp$ and $\psi_{3/2}\to Z^0\nu$ produce a gamma-ray flux with a broad
continuous energy spectrum; this flux could potentially show up in
observations of galaxy clusters or the extragalactic gamma-ray background (see
\textit{e.g.} Ref.~\cite{Bertone:2007aw, Ke:2011xw, GomezVargas:2011ph}).

Here, we briefly summarize our searches for gamma-ray signals from gravitino
dark matter in the data of the Fermi Large Area Telescope
(LAT)~\cite{Atwood:2009ez}. Firstly~\cite{Vertongen:2011mu}, we extend the
gamma-ray line analysis presented in Ref.~\cite{Abdo:2010nc} to a larger
energy range of $1$--$300\GeV$, searching for significant line signals that
might come from dark matter decay (or annihilation) in the Galactic dark
matter halo.  Secondly~\cite{Huang:2011xr}, we analyze the gamma-ray flux from
eight galaxy clusters, targets which are more sensitive to continuous spectra.
Extending previous analysis~\cite{Ackermann:2010rg, Dugger:2010ys}, we treat
the decaying dark matter signal as extended source and analyze the different
target clusters individually as well as in a combined likelihood approach.  We
present constraints on the dark matter lifetime as well as on the annihilation
cross section.  We then apply our findings to the scenario of decaying
gravitino dark matter and comment on implications for the possible observation
of long-lived superparticles at the LHC.

The remaining sections are organized as follows: In the second section, we
introduce briefly the gravitino dark matter scenario, in section three we
summarize our gamma-ray line and galaxy cluster analysis, and in the fourth
section we present the resulting limits on the gravitino lifetime and decay
width of the next-to-lightest superparticle (NLSP).

\section{Gravitino Dark Matter}
Among the different scenarios that were proposed to reconcile thermal
leptogenesis and gravitino dark matter with the standard BBN
scenario~\cite{Ellis:1984er, Kawasaki:2004qu, Pospelov:2006sc,
Kawasaki:2008qe}, a mild violation of $R$-parity that induces a rapid decay of
the NLSP before the onset of the BBN is maybe the most interesting from the
perspective of indirect dark matter searches~\cite{Buchmuller:2007ui}.  If
$R$-parity is violated, the gravitino dark matter particle becomes unstable
and subject to decay, which opens the possibility to look for its decay
products in the cosmic-ray fluxes. 

Here, we consider the supersymmetric standard model with explicit bilinear
$R$-parity violation as described in Ref.~\cite{Bobrovskyi:2010ps}.  Trading
the mass mixing parameters for $R$-parity breaking Yukawa couplings as
proposed in Ref.\,\cite{Bobrovskyi:2010ps}, the gravitino decay is a function
of a single dimensionless parameter $\zeta$, which also enters the decay width
of the NLSP.  As boundary conditions for the supersymmetry breaking parameters
of the MSSM at the grand unification (GUT) scale, we consider equal scalar and
gaugino masses $m_0 = m_{1/2}$, a zero trilinear scalar coupling $a_0=0$, and
$\tan \beta =10$. In this case, the bino-like neutralino $\widetilde \chi_1^0$
is the NLSP; the universal gaugino mass $m_{1/2}$ remains as the only
independent variable, and the gaugino masses $M_{1,2,3}$ satisfy the following
relations at the electroweak scale: $M_3/M_1 \simeq 5.9$ and $M_2/M_1 \simeq
1.9$. Electroweak precision tests (EWPT)~\cite{Buchmuller:2008vw,
Nakamura:2010zzi, Bobrovskyi:2010ps}, as well as the possible overproduction
of gravitinos in presence of the high reheating temperatures required by
standard thermal leptogenesis~\cite{Buchmuller:2004nz}, yield further bounds
on the gravitino mass like $m_{3/2} \gtrsim 30$\,GeV, and on the NLSP
neutralino mass like $100\GeV\lesssim\mN\lesssim690\GeV$ (for details see
Ref.~\cite{Vertongen:2011mu}).

\begin{figure}[t]
  \begin{center}
    \includegraphics[width=0.7\linewidth]{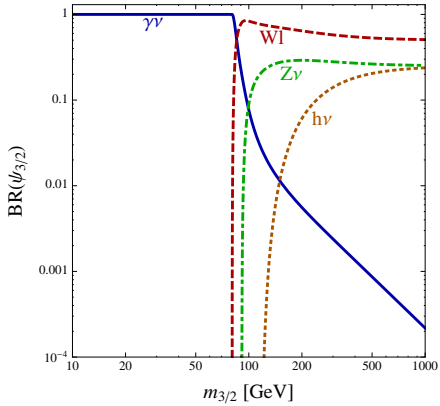}
    \vspace{-0.3cm}
  \end{center}
  \caption{Two-body decay branching ratios of the gravitino, from
  Ref.~\cite{Huang:2011xr}.}
  \label{fig:BR_spectra}
\end{figure}

The gravitino inverse decay rate into photon/neutrino pairs is given by
\cite{Takayama:2000uz, Bobrovskyi:2010ps}
\begin{align}
  \label{eq:Gdecayrate}
  \Gamma^{-1}_{\psi_{3/2}\to\gamma \nu} \simeq \frac{32 \sqrt{2}}{\alpha \zeta^2}
  \frac{\GF
  \MP^2}{\mG^3} \frac{M_1^2 M_2^2}{(M_2-M_1)^2}\;,
\end{align} 
where $\alpha$ is the electromagnetic fine structure constant, $\MP =
2.4\times 10^{18}$\,GeV the reduced Planck mass, and $\GF = 1.16\times
10^{-5}$\,GeV$^{-2}$ is the Fermi constant. The branching ratios into other
channels are presented in Fig.\ref{fig:BR_spectra}.

\section{Fermi LAT Limits}

\subsection{Gamma-Ray Lines}

\begin{figure}[t]
  \centering
  \includegraphics[width=\linewidth]{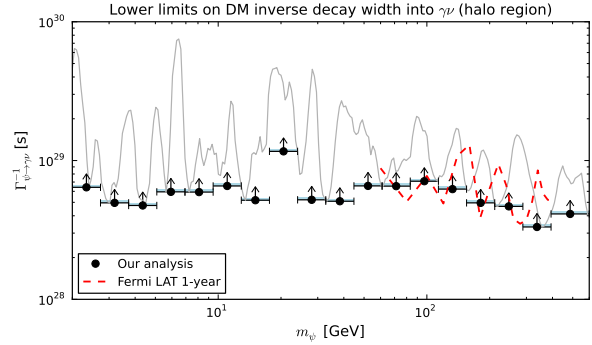}
  \caption{Lower bounds on the dark matter inverse decay width into
  monochromatic photons and neutrinos as a function of the dark matter mass
  $m_\psi$, \textit{cp.}~Ref.~\cite{Vertongen:2011mu}. The gray-solid line
  shows the $95\%$ C.L.~limits as function of the gamma-ray line energy, the
  black dots show the weakest limits obtained in certain adopted energy bands.
  The previous Fermi LAT limits from Ref.~\cite{Abdo:2010nc} are shown for
  comparison.}
  \label{fig:linelimits}
\end{figure}

Our line search is based on the measurements of the cosmic gamma-ray flux
performed by the Large Area Telescope (LAT).  The gamma-ray events that enter
our analysis are selected from the `DataClean' event class measured between 4
Aug 2008 and 17 Nov 2010.  We consider energies between 1\,GeV and 300\,GeV,
and apply the zenith angle criterion $\theta<105^\circ$ in order to avoid
contamination by the Earth's Albedo.  The expected shape of the measured
gamma-ray line spectrum is inferred from the Fermi LAT instrument response
function. In our analysis, we consider only Galactic contributions to the dark
matter signal, and take the Navarro-Frenk-White (NFW)
profile~\cite{Navarro:1996gj, Abdo:2010nc} as a reference for the dark matter
distribution (Einasto or isothermal profiles would lead to very similar
results). All profiles are normalized to $\rho_\text{dm}=0.4\GeV\cm^{-3}$ at
Sun's position. For decaying dark matter signals we choose to consider the
whole sky excluding only the Galactic disk at $|b|\leq 10^\circ$ with its
large foregrounds, since this large region features the best signal-to-noise
ratio.

The profile likelihood method~\cite{Rolke:2004mj} is used to calculate the
significance of a potential gamma-ray line contribution to the observed
gamma-ray flux. The data are modeled by a simple power law plus a line signal
at fixed energy $E_\gamma$.  Since the power law is only locally a good
approximation to the background fluxes, we use a small sliding energy window
in the fitting procedure. The size of this energy window varies between
$\pm2\sigma_{\Delta E}^{68\%}$ at low gamma-ray line energies $E_\gamma$, and
roughly $\frac13E_\gamma$ to $3 E_\gamma$ at high gamma-ray line energies.
Lifetime upper limits at the $95\%$ C.L.~are derived by increasing the line
signal and refitting the remaining parameters until the -2log(likelihood) of
the fit increases by 4 from its best-fit value.

No gamma-ray lines with $5\sigma$ significance were found in our analysis; the
corresponding limits on dark matter decay into monochromatic photons are shown
in Fig.~\ref{fig:linelimits}.

\subsection{Galaxy Cluster Observations}
\begin{figure}[t]
  \centering
  \includegraphics[width=0.9\linewidth]{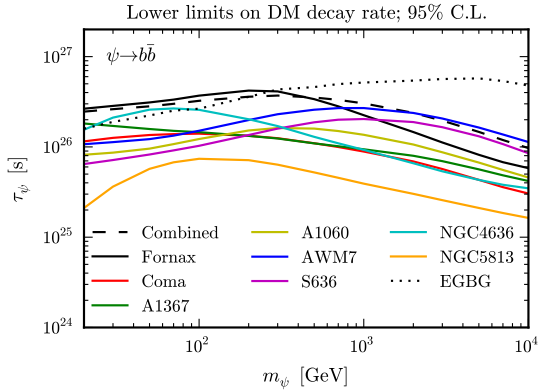}
  \caption{Lower limits on dark matter lifetime for decay into $b\bar{b}$
  final states, as function of the dark matter mass,
  \textit{cp.}~Ref.~\cite{Huang:2011xr}. Solid lines show individual cluster
  limits, the dashed line the limits from the combined likelihood analysis.
  The dotted line shows for comparison the limit that can be derived from the
  EGBG.}
  \label{fig:bbar}
\end{figure}

The eight galaxy clusters that we consider in this work are Fornax, Coma,
A1367, A1060, AWM7, S636, NGC4636 and NGC5813.  They are selected from the
extended HIFLUGCS X-ray catalog~\cite{Reiprich:2001zv, Chen:2007sz} in order
to yield large signals from dark matter decay.  The gamma-ray events entering
our analysis are taken from the \texttt{P7SOURCE\_V6} event class of the
Fermi LAT data measured between 4 Aug 2008 and 21 Jul 2011. From all events
recorded by the Fermi LAT, we select those with energies between 400\,MeV and
100\,GeV and apply the zenith angle criterion $\theta<100^\circ$ in order to
avoid contamination by the Earth's Albedo. For each galaxy cluster, we
consider photons events in a $10^\circ\times10^\circ$ squared region centered
on the cluster position.  These events are binned into a cube of
$0.1^\circ\times0.1^\circ$ pixels with 24 logarithmic energy bins.

We assume that the smooth component of the dark matter halo follows a
Navarro-Frenk-White (NFW) profile~\cite{Navarro:1996gj, Abdo:2010nc}, where
the scale radius $r_s$ and the density normalization $\rho_s$ have to be
determined from observations. We adopt the observationally obtained
concentration-mass relation from Ref.~\cite{Buote:2006kx} and use the cluster
masses derived from ROSAT PSPC X-ray observations in the extended HIFLUGCS
catalog~\cite{Chen:2007sz} to calculate the signal surface densities.  Gamma
rays from inverse Compton scattering between the CMB and the electrons and
positrons that are produced in the dark matter decay are fully taken into
account.

As above, we use the profile likelihood method to fit the data with background
and signal fluxes~\cite{Rolke:2004mj}.  For the diffuse background fluxes we
take the isotropic emission and the galactic foreground model templates
currently advocated by the Fermi LAT collaboration for point source analysis
(\texttt{iso\_p7v6source} and \texttt{gal\_2yearp7v6\_v0}).  On top of the
diffuse templates, we add the point sources from the second Fermi LAT catalog
2FGL~\cite{Collaboration:2011bm} within a radius of $12^\circ$ around the
cluster centers, as well as the extended dark matter signal. We account for
uncertainties of the cluster masses as determined by X-ray observations as a
systematic errors in our analysis. To this end, we approximate the posterior
probability for the cluster masses by log-normal distributions.  The resulting
signal uncertainties are as large as a factor two in some cases.  Finally, to
combine the statistical power of the different target regions and to reduce
the impact of the cluster mass uncertainties, we performed a combined
likelihood analysis of all eight clusters simultaneously.  In this case, the
combined likelihood function is defined as the product of the likelihood
functions for the individual clusters.  The only parameter that is bound to be
identical for all targets is the dark matter lifetime.

No significant emission from the target clusters was found. For the case of
decay into $b\bar{b}$, our resulting limits are shown in Fig.~\ref{fig:bbar}.
There, we show the limits that we obtain from the clusters individually (solid
lines), as well as the limit from the combined analysis (dashed line). The
dotted line shows limits derived from the extra-galactic gamma-ray background
(EGBG) as measured by Fermi LAT in Ref.~\cite{Abdo:2010nz} (see
Ref.~\cite{Huang:2011xr} for details).

\section{Discussion}

\begin{figure}[t]
  \begin{center}
    \includegraphics[width=\linewidth]{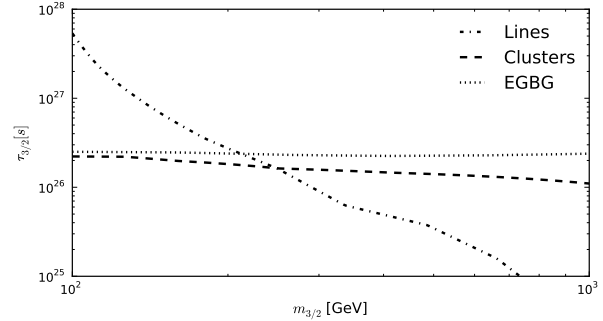}
    \vspace{-0.3cm}
  \end{center}
  \caption{Lower limits on the gravitino lifetime,
  \textit{cp.}~Ref.~\cite{Huang:2011xr}. The dot-dashed line shows the
  gamma-ray line limits, the dashed line the limits resulting from the
  combined cluster analysis (using the branching ratios shown in
  Fig.~\ref{fig:BR_spectra}), and the dotted line the EGBG limits.}
  \label{fig:bounds_tau}
\end{figure}

\begin{figure}[t]
  \begin{center}
    \includegraphics[width=\linewidth]{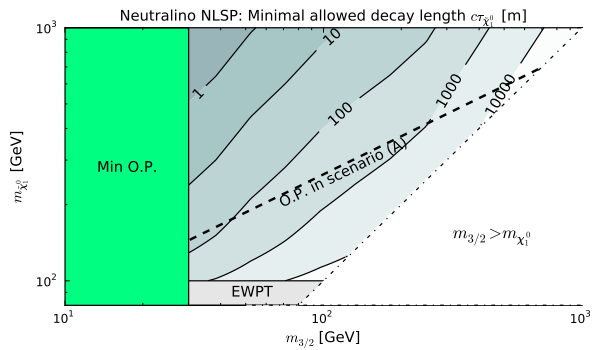}
    \vspace{-0.3cm}
  \end{center}
  \caption{Contour plot of lower bounds on the neutralino NLSP decay length,
  coming from both gamma-ray line and cluster constraints on the gravitino
  lifetime, as function of the neutralino and gravitino masses, $m_{\chi_1^0}$
  and $m_{3/2}$ respectively, \textit{cp.}~Ref.~\cite{Huang:2011xr}. The lower
  gray region is excluded by electroweak precision tests (EWPT).  For thermal
  leptogenesis, overproduction (O.P.) of gravitinos excludes at minimum the
  left green region, a limit which strengthens to the black-dashed line when
  assuming the discussed universal boundary conditions.}
  \label{fig:Neutralinodecaylength}
\end{figure}

In Fig.~\ref{fig:bounds_tau} we finally summarize the limits on the gravitino
lifetime that we obtain from our gamma-ray line searches (dotdashed) and the
galaxy cluster analysis (dashed). As one can see from this figure, while the
search for gamma-ray lines is efficient for gravitino masses $m_{3/2} \lesssim
200$\,GeV, constraints from galaxy clusters observations dominates for
$m_{3/2} \gtrsim 200$\,GeV. However, at high gravitino masses even stronger
limits come from the measured EGBG (dotted).

A neutralino NLSP heavier than 100\,GeV dominantly decays into $W^\pm\ell^\mp$
and $Z^0\nu$. The corresponding decay width is directly proportional to
$\zeta$ squared, which also enters the gravitino decay width
Eq.~\eqref{eq:Gdecayrate}. As a consequence, the two quantities can be
directly related~\cite{Bobrovskyi:2010ps}. Using the above gaugino mass
relation, lower bounds on the neutralino decay length $c\tau_{\tilde
\chi_1^0}$ can be derived. Our results are summarized in
Fig.~\ref{fig:Neutralinodecaylength}. For the parameter space allowed by EWPT
and overproduction bounds, we obtain minimal decay lengths ${\cal
O}(100\,\text{m}-100$\,km); decay lengths as small as $c\tau_{\tilde \chi_1^0}
\simeq 60$\,m are allowed for $\mG \simeq 30$\,GeV at $\mN \simeq 140$\,GeV.
Most interestingly, even these large decay lengths are in the range of
detectability of the LHC~\cite{Ishiwata:2008tp}, since processes like
$\tilde\chi^0_1\to Z^0\nu\to \mu^+\mu^-\nu$ with displaced vertices inside the
collider would lead to essentially background free
signatures~\cite{Bobrovskyi:2011vx}.

\begin{acknowledgments}
  CW thanks the organizers of the 2011 Fermi Symposium for an inspiring
  conference and the Kavli Institute for Theoretical Physics China, Beijing,
  for kind hospitality.
\end{acknowledgments}





\bibliography{}
\end{document}